\documentstyle[twoside,fleqn,espcrc2,epsf,graphicx]{article}

\title{Lattice Study of the Massive Schwinger Model with a $\theta$ term 
under L\"uscher's "Admissibility" condition\thanks{Talk presented by
H. Fukaya.}}

\author{
H.~Fukaya
\address[Ox]{Yukawa Institute for Theoretical Physics, Kyoto
University, Kyoto 606-8502, Japan} 
and T.~Onogi\addressmark[Ox]\thanks{Supported by 
Grant-in-Aid for Scientific Research on Priority Areas,  No.13135213.}}

\begin{document}

\begin{abstract}
We present a numerical study of the massive two-flavor QED in two 
dimensions with the gauge action proposed by L\"uscher, which allows 
only ``admissible'' gauge fields. 
We find that the admissibility condition does not allow any
topology changes  by the local updation in Hybrid Monte Carlo
algorithm so that the configurations in each topological sector 
can be generated separately. By developing a new method to sum 
over different topological sectors, we investigate $\theta$ vacuum 
effects. Combining with domain-wall fermion action, we 
obtain the fermion mass dependence and $\theta$ dependence of the 
meson masses, which are consistent with the analytic results by 
mass perturbation in the continuum theory.
\vspace{0mm}
\end{abstract}

\maketitle
\section{Introduction}

The studies of chiral symmetries on the lattice 
have made much progress recently.
There are two keys in these developments;
(1) Ginsparg-Wilson relation \cite{Ginsparg:1981bj}:
\begin{equation}
\gamma_{5}D+D\gamma_{5}=aD\gamma_{5}D,
\end{equation}
and 
(2) L\"uscher's ``admissibility'' condition 
\cite{Luscher:1998du,Luscher:1999un}:
\begin{equation}\label{admi}
\|1-U^{P}_{\mu\nu}(x)\| \leq \epsilon\;\;\;
\mbox{for all}\;\;x,\mu,\nu.
\end{equation}
where $a$ denotes the lattice spacing, $U^{P}_{\mu\nu}(x)$
denotes a plaquette and $\epsilon$ is a fixed constant.

The first key, the Ginsparg-Wilson relation, 
gives a redefinition of chiral symmetries
without fermion doublers at the classical level.
The second key, the "admissibility" condition Eq.~(\ref{admi}), 
makes gauge fields smooth. Under this condition 
the gauge fields are classified by a topological charge which 
corresponds to that of the continuum theory, giving a well-defined
chiral anomaly at the quantum level.

In two-dimensional QED, the topological charge is defined as
\begin{eqnarray}
Q &= \frac{1}{2\pi}\sum_{x,\mu,\nu} \left[-\frac{i}{2}\epsilon_{\mu\nu}
\ln U^{P}_{\mu\nu}(x)\right]\;\;\mbox{(integer)}\nonumber\\
&\to_{a\to 0}
\frac{1}{2\pi}\int_{T^{2}} d^{2}x\frac{1}{2}\epsilon_{\mu\nu}
F_{\mu\nu}(x).
\end{eqnarray}

While numerical studies on Ginsparg-Wilson fermion
have been performed,
no study of L\"uscher's  admissible gauge fields
has been done before.

We present a numerical study of two-dimensional QED with 
L\"uscher's gauge action which satisfies admissibility condition
automatically \cite{Fukaya:2003ph}. We show that topological structure
is realized on the lattice, $\theta$ dependence of observables
can be evaluated by our new method. As a result we obtain 
the mass of isotriplet meson which is consistent with the continuum theory.
This work may offer a new approach to investigate ``topology''
in the lattice gauge theories.
\section{Lattice Simulations}
We take L\"uscher's gauge action combined with the domain-wall
fermion action;
\begin{eqnarray}
S &=& \beta S_{G} + \sum_{\mbox{\small flavor}=1}^{2}S_{F},\\
S_{G}&=& \left\{\begin{array}{l}
\displaystyle{\sum_{x,\mu,\nu}}\frac{(1-{\rm Re}U^{P}_{\mu\nu}(x))}
{1-(1-{\rm Re}U^{P}_{\mu\nu}(x))/\epsilon}
\\  \hspace{1in}\mbox{if admissible}
\\ \infty \hspace{1in} \mbox{otherwise}
\end{array}\right. ,\label{eq:admaction2}\\
S_{F} &:& \mbox{the domain-wall fermion action}
\nonumber\\ &&\mbox{including Pauli-Villars fields},
\end{eqnarray}
where $\beta =1/e^{2}$.
The L\"uscher's condition is automatically satisfied
with this action.

We use the Hybrid Monte Carlo (HMC) method with pseudo-fermions and
conjugate gradient (CG) algorithm for the inversion of the fermion 
matrix. 
50 molecular dynamics steps with step-size $\Delta \tau=0.02$
are performed for each trajectory and configurations are updated 
per 10 trajectories.
We also check the admissibility at the Metropolis test.
The simulations are carried out on a $16\times 16\times 6$ lattice
with $\beta=0.5$ and $\epsilon=1.0$.
Fermion mass $m$ is taken to be 0.1,0.2,0.3 and 0.4.

\section{Topological Sectors and Reweighting}

\begin{figure}[thb]
\epsfxsize=\hsize
\includegraphics[height=7cm,angle=-90]{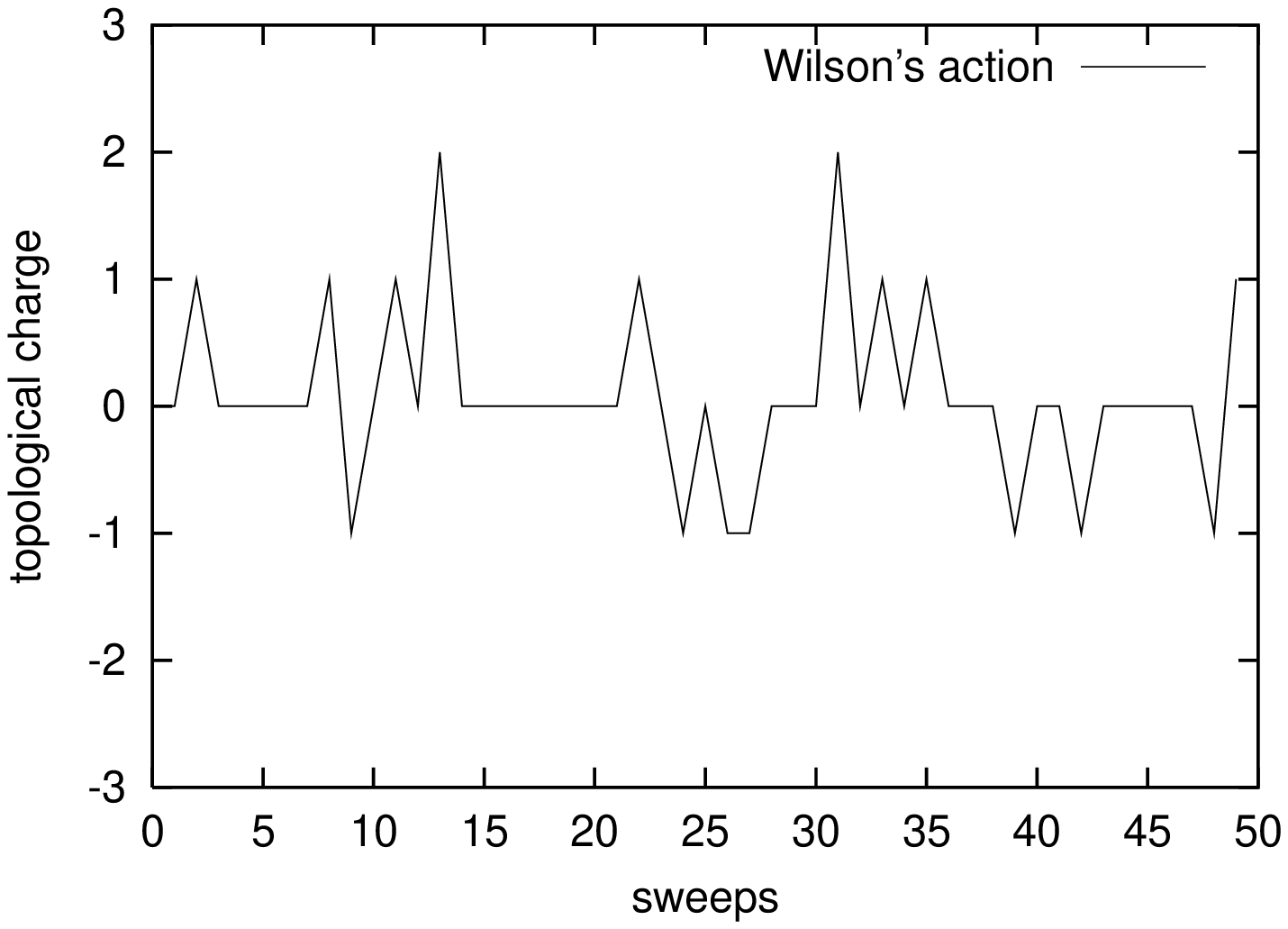}
\epsfxsize=\hsize
\includegraphics[height=7cm,angle=-90]{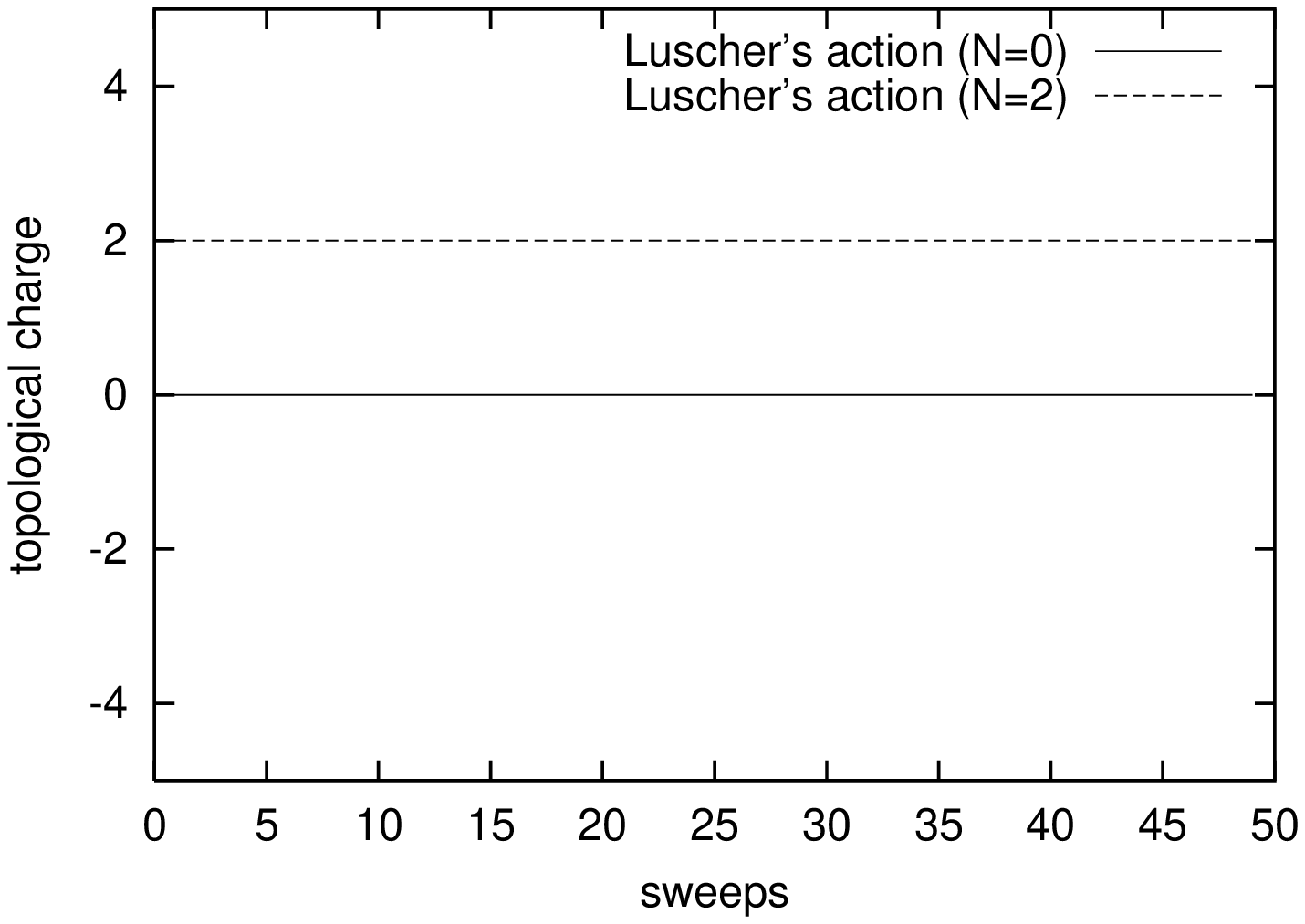}
\caption{The comparison of the evolution of the
topological charge with Wilson's gauge action and L\"uscher's gauge
action for the same lattice spacings determined from the string tension.}
\label{fig:top}
\vspace{-15mm}
\end{figure}

We set the initial link variables as follows,
\begin{eqnarray}\label{eq:const background}
U^{cl[N]}_{1}(x,y) &=& \exp\left\{
-\frac{2\pi Ni}{L}
\delta_{x,L} y \right\},
\nonumber\\
U^{cl[N]}_{2}(x,y) &=& \exp\left\{
\frac{2\pi Ni}{L^{2}} x\right\},
\end{eqnarray}
where $L=16$ is the lattice size. 
This configuration gives constant background
electric fields over the torus and minimizes the 
gauge action in the sector with topological charge $N$.
As Fig.\ref{fig:top} shows, L\"uscher's action generates 
configurations without any topology changes.

Thus  admissible gauge fields are separated into topological sectors
and configurations are generated in each sector with L\"uscher's action.
The next task is to sum up the expectation values in each sector
with correct weights.
\begin{equation}
\langle  O \rangle^{full}_{\beta,m} = 
\frac{ \sum_{N=-\infty}^{+\infty}e^{iN \theta}
       \langle  O \rangle^{N}_{\beta,m}
       R^{N}(\beta,m) }
     { \sum_{N=-\infty}^{+\infty}e^{iN \theta}
       R^{N}(\beta,m) },
\label{O_full}
\end{equation}
where
\begin{eqnarray*}
R^{N}(\beta,m)=\frac{Z_{N}(\beta,m)}{Z_{0}(\beta,m)},\hspace{1in}
\nonumber\\
\langle  O \rangle^{N}_{\beta,m} : \mbox{expectation value in each sector} 
\hspace{0.4in}\\
Z_N(\beta,m) : \mbox{generating functional in each sector.} 
\end{eqnarray*}
We developed a new method to calculate the reweighting factor
$R^{N}(\beta,m)$ \cite{Fukaya:2003ph}. By separating $R^{N}(\beta,m)$ 
into three factors as
\begin{eqnarray}
R^{N}(\beta,m)=\exp(-\beta S^{N}_{G\;min})\hspace{0.5in}
\nonumber\\
\times
\frac{\int d\nu_1 d\nu_2 \det(D^{N}_{DW})^{2}/\det(D^{N}_{PV})^{2}}{\int d\nu_1 d\nu_2 \det(D^{0}_{DW})^{2}/\det(D^{0}_{PV})^{2}}
\nonumber\\
\times e^{\left[\int_{\beta}^{\infty}d\beta^{\prime}
\left(\langle  S_{G} - S^{N}_{G\;min}\rangle^{N}_{\beta^{\prime},m}
-\langle  S_{G} \rangle^{0}_{\beta^{\prime},m}\right)\right]},
\end{eqnarray}
where $S^{N}_{G min}$ in the first factor denotes 
the minimum of the gauge action in $N$ sector, 
the second factor is the contribution from free fermion determinants
given by the integral of Polyakov loops , 
and the third one is a $\beta$ integral of the expectation value of the 
gauge action.

Once the reweighting factor is known,
we can evaluate theta dependence of observables as in Eq.~(\ref{O_full}).

\section{Meson Masses}

We calculated the isotriplet meson propagators at various $\theta$ as
\begin{eqnarray}
\sum_{y}\langle \pi(x,y)\pi(0,0)\rangle_{full}=\hspace{1in}
\nonumber\\
\sum_{N=-4}^{4}e^{iN\theta}
\sum_{y}\langle \pi(x,y)\pi(0,0)\rangle^{N}_{\beta,m}
R^{N}(\beta,m).
\end{eqnarray}
Here we have ignored $|N| > 4$ sectors since they only give 
contributions less than 1.2 \% of zero sector.
Fitting them into hyperbolic cosine function,
we evaluated the meson mass.

\begin{figure}[htb]
\begin{center}
\includegraphics[height=7cm,angle=-90]{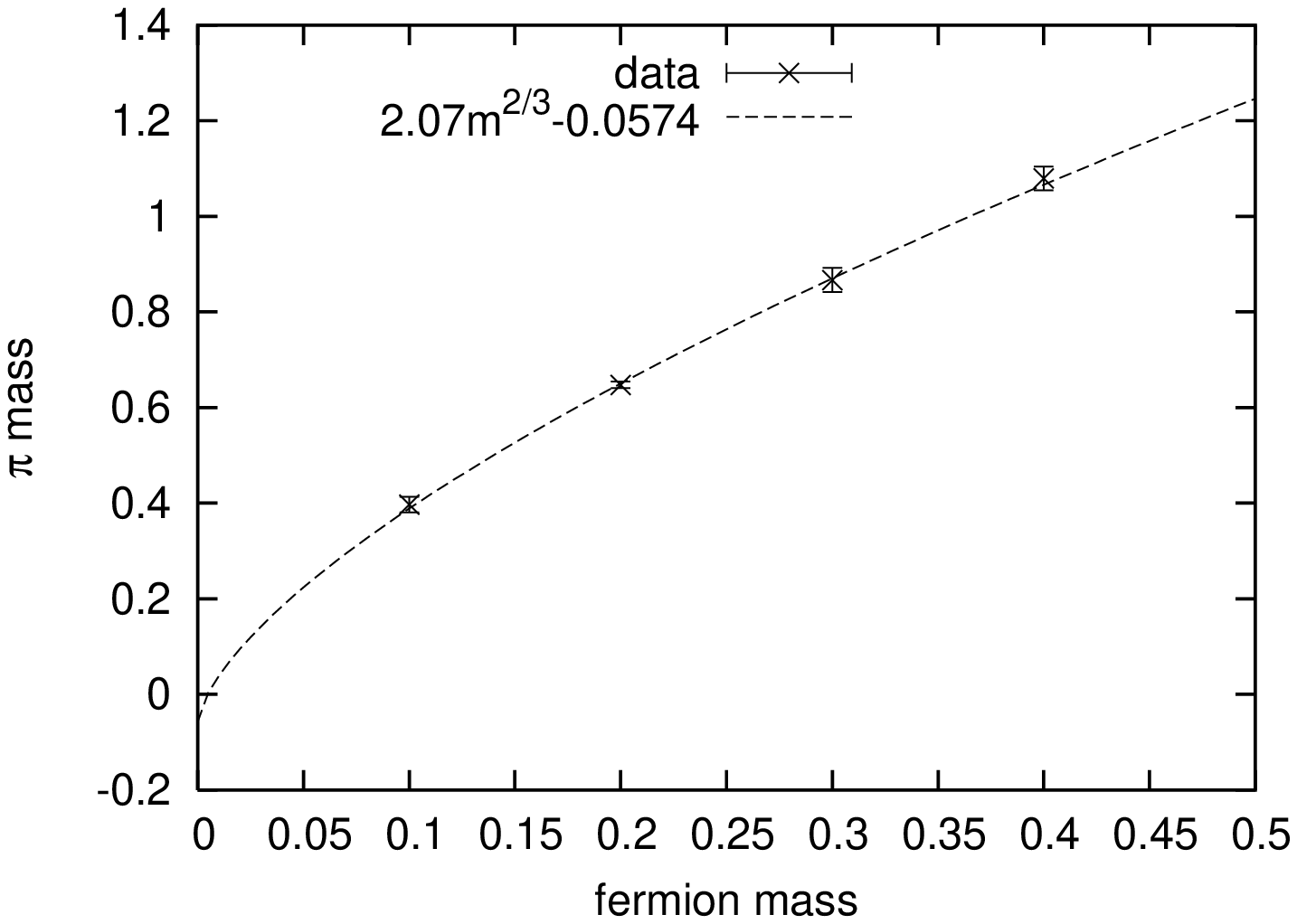}
\vspace{-7mm}
\caption{The fermion mass dependence of the pion mass at $\theta=0$.
}\label{fig:pimass}
\includegraphics[height=7cm,angle=-90]{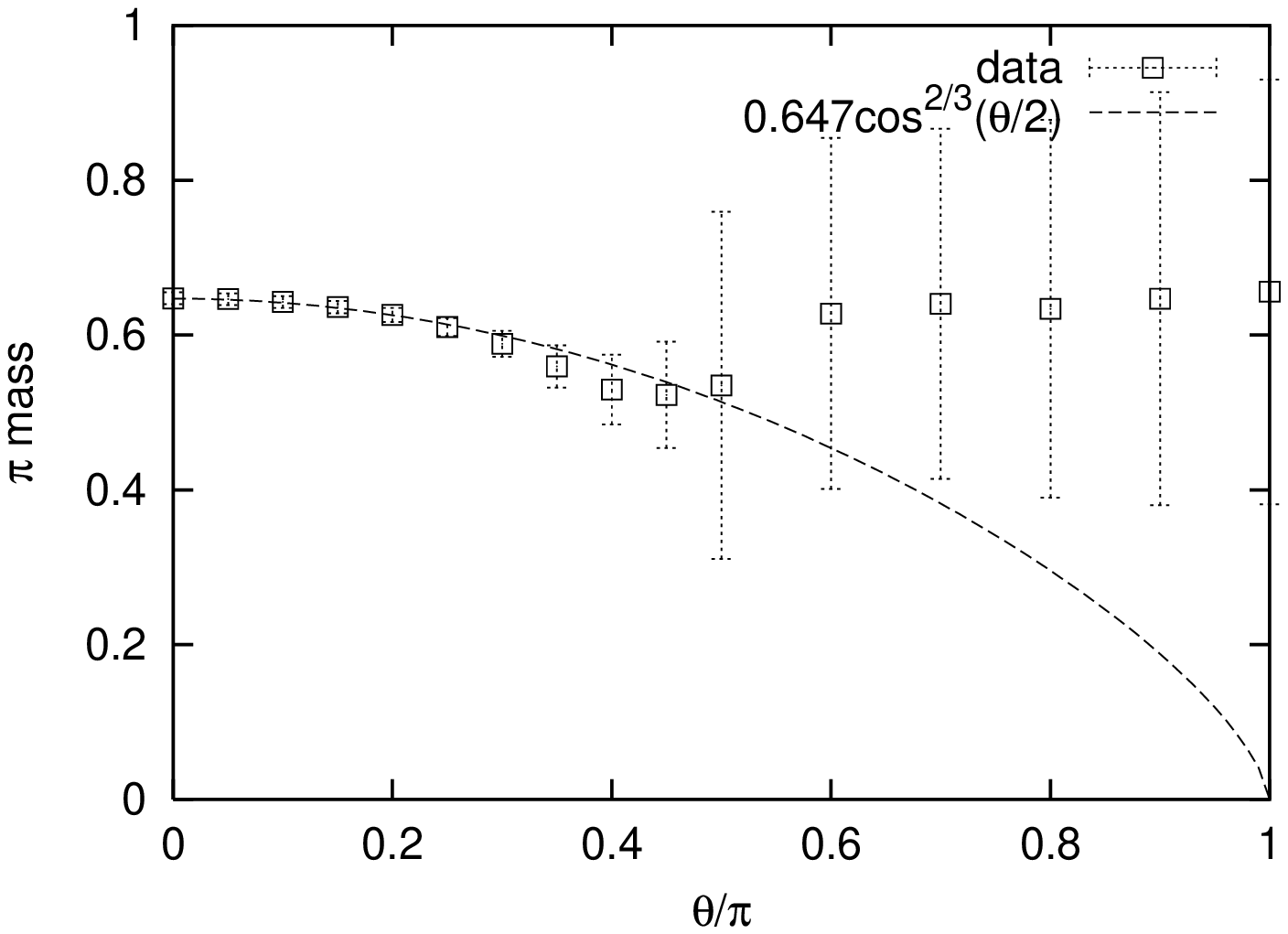}
\vspace{-7mm}
\caption{$\theta$ dependence of the pion mass at
 $m=0.2$.}\label{fig:pitheta}
\end{center}
\vspace{-7mm}
\end{figure}

As Figs.\ref{fig:pimass} and \ref{fig:pitheta} show, 
the fermion mass and $\theta$ dependence of the pion mass are 
consistent with that of continuum theory at small $\theta$
\cite{Coleman:1975pw,Coleman:1976uz,Frohlich:mt};
      \begin{equation}
       m_{\pi} \propto 
	(m\cos\frac{\theta}{2})^{\frac{2}{3}}.
      \end{equation}

Moreover, it is remarkable the residual mass in the chiral 
limit is consistent with zero within statistical error 
even in this strong coupling regime; 
\begin{equation}
m_{\pi}(m\to 0)=-0.057\pm0.060.
\end{equation}

\section{Summary}

We performed a numerical simulation of 2 dimensional theory
with L\"uscher's gauge action combined with domain-wall
fermion action.
We also proposed a new method for summing over
different topological sectors.

Applying L\"uscher's condition to numerical simulations 
we find that the lattice gauge fields are separated into 
topological sectors. Our reweighting method enables the summation 
over topological sectors. We obtained the results which are 
qualitatively consistent with that in the continuum theory.
It is also notable that the residual mass of the domain-wall 
fermion in the chiral limit is consistent with zero even in 
the strong coupling regime.

As future works, we plan to perform more precise and quantitative 
studies of 2 dimensional QED and comparison to 
the results in conventional lattice formalism \cite{Schwinger}.
Studying how to extend our approach to 4 dimensional 
QCD is also important.

\end{document}